\documentclass[sigconf,natbib=true,anonymous=false,review=false]{acmart}
\usepackage{multicol}
\usepackage{multirow}
\usepackage{amsfonts}
\usepackage{bm}
\usepackage{amsmath}
\usepackage{amsthm}
\newcommand{\x}{\bm{x}}

\AtBeginDocument{%
  \providecommand\BibTeX{{%
    \normalfont B\kern-0.5em{\scshape i\kern-0.25em b}\kern-0.8em\TeX}}}

\copyrightyear{2024} 
\acmYear{2024} 
\setcopyright{acmlicensed}\acmConference[CIKM '24]{Proceedings of the 33rd ACM International Conference on Information and Knowledge Management}{October 21--25, 2024}{Boise, ID, USA}
\acmBooktitle{Proceedings of the 33rd ACM International Conference on Information and Knowledge Management (CIKM '24), October 21--25, 2024, Boise, ID, USA}
\acmDOI{10.1145/3627673.3679534}
\acmISBN{979-8-4007-0436-9/24/10}

\begin{document}

\title{Aligning Query Representation with Rewritten Query and Relevance Judgments in Conversational Search}

\author{Fengran Mo}
\orcid{0000-0002-0838-6994}
\affiliation{%
  \institution{RALI, Université de Montréal}
  \city{Montréal}
  \state{Québec}
  \country{Canada}
}
\email{fengran.mo@umontreal.ca}

\author{Chen Qu}
\orcid{0000-0002-3273-7109}
\affiliation{%
  \institution{University of Massachusetts Amherst \city{Amherst}
  \state{MA}
  \country{USA}}
}
\email{mail@cqu.org}

\author{Kelong Mao}
\orcid{0000-0002-5648-568X}
\affiliation{%
  \institution{Renmin University of China} 
  \city{Beijing}
  \country{China}}
\email{mkl@ruc.edu.cn}

\author{Yihong Wu}
\orcid{0009-0009-2680-4107}
\affiliation{%
  \institution{RALI, Université de Montréal}
  \city{Montréal}
  \country{Canada}}
\email{yihong.wu@umontreal.ca}

\author{Zhan Su}
\orcid{0000-0001-5189-9165}
\affiliation{%
  \institution{University of Copenhagen}
  \city{Copenhagen}
  \country{Denmark}
}
\email{zhan.su@di.ku.dk}

\author{Kaiyu Huang}
\orcid{0000-0001-6779-1810}
\affiliation{%
  \institution{Beijing Jiaotong University}
  \city{Beijing}
  \country{China}}
\email{kyhuang@bjtu.edu.cn}

\author{Jian-Yun Nie}
\orcid{0000-0003-1556-3335}
\affiliation{%
  \institution{RALI, Université de Montréal}
  \city{Montréal}
  \state{Québec}
  \country{Canada}
}
\email{nie@iro.umontreal.ca}

\renewcommand{\shortauthors}{Fengran Mo et al.}

\begin{abstract}
Conversational search supports multi-turn user-system interactions to solve complex information needs. Different from the traditional single-turn ad-hoc search, conversational search encounters a more challenging problem of context-dependent query understanding with the lengthy and long-tail conversational history context. 
While conversational query rewriting methods leverage explicit rewritten queries to train a rewriting model to transform the context-dependent query into a stand-stone search query, this is usually done without considering the quality of search results.
Conversational dense retrieval methods use fine-tuning to improve a pre-trained ad-hoc query encoder, but they are limited by the conversational search data available for training. 
In this paper, we leverage both rewritten queries and relevance judgments in the conversational search data to train a better query representation model. The key idea is to align the query representation with those of rewritten queries and relevant documents. The proposed model -- \textbf{Q}uery \textbf{R}epresentation \textbf{A}lignment \textbf{C}onversational \textbf{D}ense \textbf{R}etriever, QRACDR, is tested on eight datasets, including various settings in conversational search and ad-hoc search. The results demonstrate the strong performance of QRACDR compared with state-of-the-art methods, and confirm the effectiveness of representation alignment.
\end{abstract}





\maketitle

\section{Introduction}
Conversational search enables users to interact with the system in a multi-turn fashion to satisfy their complex information needs. It is envisioned to be the interaction mode for next-generation search engines~\cite{gao2022neural}. Compared to the traditional single-turn ad-hoc search scenario, the lengthy and long-tail historical context in conversational search raises the difficulty for search intent understanding in context-dependent conversation turns.

To capture the real information needs in each query turn, an intuitive method is Conversational Query Rewriting (CQR)~\cite{elgohary2019can}, which leverages a rewriting model to transform the context-dependent queries into stand-alone ones, which can then be used by any ad-hoc search models for retrieval. Nevertheless, optimizing the rewriting model for the downstream search task in a two-stage, \textit{rewrite-then-search}, approach is challenging. This difficulty arises due to the separate generation process in rewriting, which disrupts the back-propagation of gradients, resulting in sub-optimal performance~\cite{wu2022conqrr,mo2023convgqr,mao2023search}.
Another typical method is Conversational Dense retrieval (CDR)~\cite{yu2021few}, which tries to learn, in an end-to-end manner, a conversational dense retriever to encode the user's real search intent and candidate documents into a learned embedding space. The implicit query understanding ability is expected to be improved via fine-tuning with document relevance signals in conversational search sessions. 

\begin{figure*}[t]
    \centering  \includegraphics[width=0.95\textwidth]{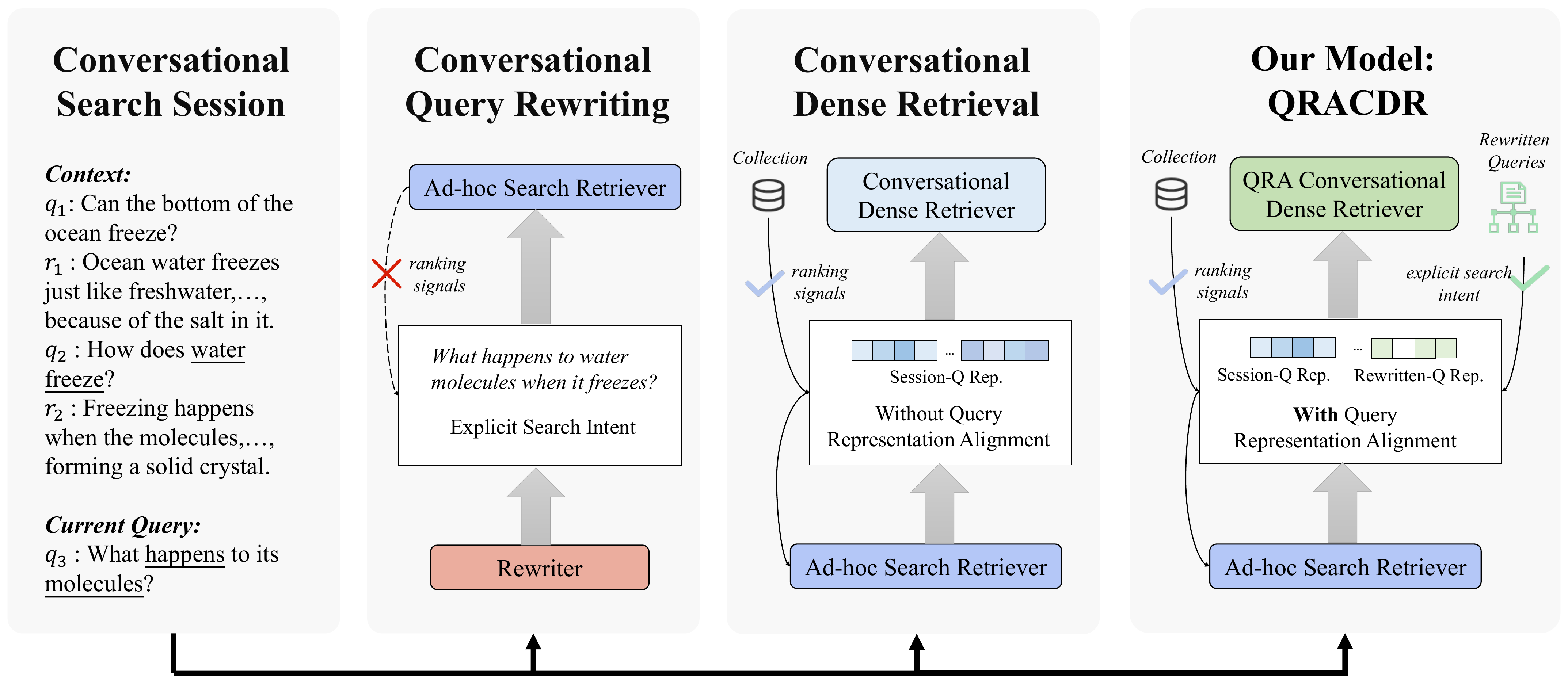}
    \caption{A conceptual illustration for the three types of methods. QRACDR has a Query Representation Alignment goal to help achieve more effective end-to-end optimization toward search within an ongoing conversation.}
    \label{fig: comparison}
\vspace{-2ex}
\end{figure*}

However, most of the existing conversational dense retrieval models~\cite{yu2021few,lin2021contextualized,kim2022saving,mo2024history} are simply learned by fine-tuning the pre-trained ad-hoc encoders via contrastive learning on conversational search data~\cite{anantha2021open,adlakha2022topiocqa}. This is because 1) the initial ad-hoc encoders undergo pre-training exclusively using concise ad-hoc queries; and 2) the current conversational search data, primarily human-generated, is not as abundant as ad-hoc search data due to the limited deployment of real conversational search systems. Consequently, the straightforward fine-tuning approach lacks sufficient supervision information to effectively adapt ad-hoc encoders to the intricate and noisy conversational search setting.
This problem is made even worse by the limited relevance judgments available in conversational search data: only a single relevant document per query is provided in the datasets (i.e., the supervision signals are weak.).

Apart from the aforementioned issues, most existing studies only leverage either the rewritten query in CQR or the relevance judgment between query turn and document in CDR as supervision signals, rather than both together. 
\citet{yu2021few} attempt to incorporate manual queries by knowledge distillation within dense retrieval (ConvDR). Although this work considers the alignment between the conversation session and the rewritten query, it does not deliberate on the alignment between the session and the gold document (nor the misalignment with the negative documents). As we further reveal in Section~\ref{sec: theory}, this gap can result in a fundamental degradation in alignment power.
 
To better understand the user search intent within the ongoing conversation, we leverage available rewritten queries used/produced by the CQR and relevance judgments in training data to learn a better model through query representation alignment.
Figure~\ref{fig: comparison} illustrates the differences between the two existing paradigms and our method.
We expect the query encoder that leverages both the rewritten query and relevance signals to approximate the session query representation toward not only the relevant document but also the rewritten query. 
This is similar to the very assumption behind many relevance feedback approaches~\cite{tao2007exploration,lavrenko2017relevance} used in information retrieval, which tries to create a query representation from the original query to become closer to relevant documents, and far away from irrelevant ones. 
In terms of conversational search, the key idea we exploit in this paper is to align a session query representation with those of the rewritten query and the relevant document in the neural representation space. The alignment losses would push the query representation closer to them. In addition, we also consider negative samples, and the query representation is pushed away from them.

The above idea is implemented in an \textbf{Q}uery \textbf{R}presentation \textbf{A}lignment \textbf{C}onversational \textbf{D}ense \textbf{R}etriever, QRACDR, with several training strategies that incorporate different alignment mechanisms.
Specifically, we first design a straightforward and effective strategy to optimize the distance between the learnable session query with both the pre-encoded rewritten query and the relevant document, which is based on the Mean Square Error (MSE) paradigm. We theoretically define the ``alignment'' area in the representation space as the region where the encoded session query is supposed to fall in, and provide preliminary experiments to confirm the usefulness of such alignments.
To further facilitate the conversational dense retriever training, we incorporate negative samples on both the alignment-originated MSE and the commonly used contrastive learning (CL) paradigm. 
Compared with previous dense retrieval models, QRACDR exploits a new principled idea for query representation.
Experimental results on eight datasets, covering usual and low-resource conversational search, and ad-hoc search scenarios, show that QRACDR can consistently outperform state-of-the-art baselines in the usual evaluation and is effective in other settings. A series of thorough analyses are conducted to understand the behind-the-scene behavior of the models.

The contributions of this paper include: (1) We propose QRACDR to leverage available rewritten queries and conversational search data to supervise the model to build better query representation for search intent via alignment, enhancing the model performance via the complementary effects of both existing CQR and CDR methods.
(2) We provide theoretical assumptions and perform empirical verification of the need for query representation alignment in conversational search and design several training strategies in QRACDR.
(3) We demonstrate the effectiveness of our QRACDR on eight datasets with different settings, where the best strategy surpasses the state-of-the-art CQR and CDR baselines. Our analyses help to understand the behind-the-scene behavior of the models.

\section{Related Work}
\label{sec: Related Work}
Conversational search is an information-seeking process through interactions with a conversation system~\cite{Zamani2022ConversationalIS}. In practice, it can be considered as a task to iteratively retrieve documents for users' queries in a multi-round dialog. Two research lines are conducted to achieve conversational search: conversational query rewriting (CQR) and conversational dense retrieval (CDR). The CQR methods try to convert the context-dependent query into a standalone query, and then apply an off-the-shelf retriever as the ad-hoc search. Existing studies try to select useful tokens from the conversation context~\cite{2020Making,voskarides2020query,fang2022open} or train a generative rewriter model with conversational sessions to mimic the human rewrites~\cite{yu2020few,lin2020conversational,vakulenko2021question}. To optimize query rewriting for the search task, some studies adopt reinforcement learning~\cite{wu2022conqrr,chen2022RLCQR} or apply the ranking signals with the rewriting model training~\cite{mo2023convgqr,mao2023search}, while others jointly learn query rewriting and context modeling~\cite{qian2022explicit}.
Besides, some recent methods are proposed to directly prompt the LLMs to generate the rewrites~\cite{jang2023itercqr,ye2023enhancing,mao2023large,mo2024chiq,mo2024leverage}.
On the other hand, conversational dense retrieval~\cite{qu2020open,mao2024chatretriever} directly encodes the whole conversational search session to perform end-to-end dense retrieval. The common practice is to employ contrastive learning with positive and negative passages. 
Although a similar ConvDR~\cite{yu2021few} study considers the alignment between the conversation session and the manual queries through knowledge distillation, it does not specifically consider the alignment between the session and the relevant document (nor the discrepancy with the negative documents) as our method does.
Besides that, existing methods also try to improve the session representation through context denoising~\cite{lin2021contextualized,mao2022curriculum,krasakis2022zeco,mo2023learning,jin2023instructor,cheng2024interpreting}, data augmentation~\cite{dai2022painting,mao2022convtrans,mo2024convsdg,chen2024generalizing}, and hard negative mining~\cite{kim2022saving,mo2024history}. 
Different from them, we combine the supervision signal from both CQR and CDR methods to achieve query representation alignment with theoretical and empirical studies.

\section{Query Representation Alignment Hypothesis in Conversational Search}
\label{sec: theory}
\subsection{Task Denfinition}
Following the literature described in Section~\ref{sec: Related Work}, we define the conversational search task as finding the relevant documents $d^+$ from a large collection $D$ to satisfy the information needs of the current query turn $q_n$ based on the given historical conversational context $\mathcal{H}=\{q_i, r_i\}_{i=1}^{n-1}$, where  $q_i$ and $r_i$ denote the $i$-th query turn and the system's response.
Compared with the ad-hoc search, the current query turn $q_n$ is context-dependent, and thus, requires the retriever to formulate a search query by considering its context.
A common practice is to reformulate the current turn into a session query $q_n^s=\mathcal{H} \circ q_n$ by concatenating with the context as the input for training the conversational dense retriever. 
Besides, a stand-alone rewritten query $q_i^{\prime}$, provided either by manual annotation or automatic generation for each turn, is helpful to obtain such understanding with the suitably designed mechanism.

\subsection{Hypothesis Verification}
Intuitively, an ideal conversational dense retriever is expected to encode the session query representation $q_n^s$ close to its relevant documents $d_n^+$. In this case, minimizing the distance between $q_n^s$ and $d_n^+$ via a single MSE loss function would be desirable. 
However, it does not work as expected in practice, due to the limitation of using only one MSE loss objective for $q_n^s$ and $d_n^+$, which only shrinks the distance between $q_n^s$ and $d_n^+$ but does not restrict the location of $q_n^s$ in the representation space. The location of $q_n^s$ affects the query representation alignment effect, which we will further explain in the following sections. 

\begin{table*}[t]
    \centering
    \caption{Preliminary experiments for query representation alignment hypothesis verification, showing the assumption of the existing of ``aligned area'' for better session query representation.}
    \vspace{-3ex}
    \begin{tabular}{l|cccc|cccc}
    \toprule
        \multicolumn{1}{c|}{\multirow{2}{*}{Representation}} & \multicolumn{4}{c|}{TopiOCQA} & \multicolumn{4}{c}{QReCC}  \\ 
        \cmidrule(lr){2-5} \cmidrule(lr){6-9} 
        & MRR & NDCG@3 & Recall@10 & Recall@100 & MRR & NDCG@3 & Recall@10 & Recall@100 \\ 
        \midrule
        Aligned ($\mathcal{V}_{\text{align})}$ & 89.7 & 90.1 & 96.5 & 99.3 & 94.1 & 92.2 & 96.5 & 98.2 \\
        Orignal ($q_n^{\prime}$)& 10.3 & 9.12 & 19.1 & 35.7 & 42.5 & 39.8 & 62.6 & 79.3 \\
        Non-aligned ($\mathcal{V}_{\text{non-align}}$) & 1.10 & 1.02 & 1.09 & 1.25 & 1.13 & 1.06 & 1.08 & 1.22\\
        \bottomrule
     \end{tabular}
     \label{table: PrelimQRAry}
\vspace{-3ex}
\end{table*}

\subsubsection{The Importance of Query Representation Alignment}
In conversational dense retrieval, two assumptions have been introduced in~\cite{yu2021few}: 1) The representation of the rewritten query $q_n^{\prime}$ and the session query $q_n^s$ should be similar because they share the same underlying information needs. 2) The representation of meaningful information in a document remains the same whether serving ad-hoc or conversational search, allowing for shared representation in both scenarios. 
The assumptions encourage better query representation alignment between the rewritten query and the session query to achieve better conversational search results.
Thus, the explicit search intent contained in the rewritten query $q_n^{\prime}$, which could be addressed by the initial ad-hoc search retriever, becomes a crucial feature to achieve the query representation alignment. 

\subsubsection{Hypothesis of Alignment Area}
Intuitively, we should encode the session query $q_n^s$ in an area around the rewritten query $q_n^{\prime}$ and relevant document $d_n^{+}$ in the representation space, i.e., close to both the rewritten query and the relevant document.
To verify our assumption, we conduct the preliminary experiments and report in Table~\ref{table: PrelimQRAry}. With the backbone ad-hoc search dense retriever ANCE~\cite{xiong2020approximate}, we first encode the relevant document and the rewritten query as $d_n^+$ and $q_n^{\prime}$ and define the aligned and non-aligned representations as Eq.~\ref{eq: assumption}. Then we directly perform retrieval with them.
From Table~\ref{table: PrelimQRAry}, we observe a huge performance gap between the representation with and without good query representation alignment. In addition, the original representation cannot directly obtain satisfactory results, indicating the necessity of conversational fine-tuning and query representation alignment.
\begin{equation}
\label{eq: assumption}
   \mathcal{V}_{\text{align}} = \left(q_n^{\prime} + d_n^+\right) / 2, \quad
   \mathcal{V}_{\text{non-align}} = \left(q_n^{\prime} - d_n^+\right) / 2
\end{equation}

From Table~\ref{table: PrelimQRAry}, we can see there is a huge performance gap between the representation with and without good query representation alignment. In addition, the original representation cannot directly obtain satisfactory results, indicating the necessity of conversational fine-tuning and query representation alignment.
The observations suggest that there should be some areas in the representation space with better query representation alignment effect when the rewritten query and relevant document are constrained and available. 
To figure out such an aligned area, we assume a hyper-sphere $\mathcal{S}_\epsilon$ with radius $\epsilon$ exists in the learned representation space $\mathcal{X} \in \mathbb{R}^m$, and $\mathcal{S}_\epsilon \subseteq \mathcal{X}$. The small error $\epsilon \in \mathbb{R}$ is defined as $||q_n^s - d_n^+||_2$. Then the hyper-sphere corresponding to the aligned area is defined as 
\begin{equation}
    \mathcal{S}_\epsilon = \{\x \in \mathbb{R}^m \mid  ||\x - d_n^+||_2=\epsilon\}
\end{equation}

Without any prior knowledge,  $q_n^s$ is assumed to be uniformly distributed on the sphere $\mathcal{S}_\epsilon$.
Since the available rewritten query $q_n^{\prime}$ is also pre-encoded, together with the observation in the preliminary experiments, there should be some areas with better query representation alignment impact on the sphere $\mathcal{S}_\epsilon$, where we called ``aligned area'' in this paper. Our goal of encoding $q_n^s$ is to make it fall in the aligned area.

\subsubsection{Aligned and Non-Align Area}
\begin{figure}
    \centering
    \includegraphics[width=0.35\textwidth]{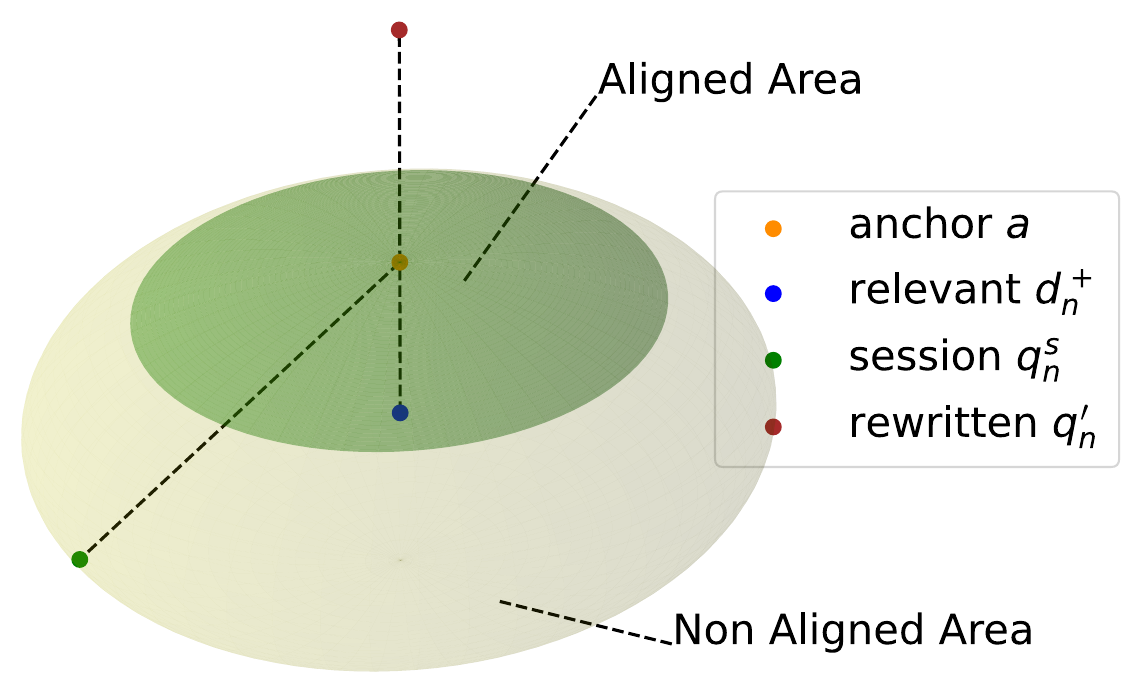}
    \caption{A conceptual illustration of the defined hyper-sphere $\mathcal{S}_\epsilon$ and the corresponding conceptions on it. The region outside the hyper-sphere denotes the whole representation space $\mathcal{X}$. The goal of achieving query representation alignment is to enable the learned session query $q_n^s$ fall into the aligned area with the help of the fixed representation of rewritten query $q_n^{\prime}$ and relevant document $d_n^+$.}
    \label{fig: sphere}
    \vspace{-4ex}
\end{figure}

In this section, we first define the aligned and non-aligned areas, and then we analyze the impact of $q_n^s$ falling in each area.
Intuitively, the aligned area on the sphere should be around the anchor $\bm{a}$ (intersection point of the line connecting $q_n^{\prime}$ and $d_n^+$ on the sphere), while the non-align area should be the remaining area. 
Their definitions are:
\begin{definition}[Anchor]
    The point $\bm{a} = 
    \epsilon\frac{q_n^{\prime} - d_n^+}{||q_n^{\prime} - d_n^+||_2} \in \mathcal{S}_\epsilon$ is defined as the anchor on the sphere.
\end{definition}
\begin{definition}[Aligned Area]
    For $\alpha \in [0,1)$, the space $\mathcal{G}_{\alpha} = \{\x \in \mathcal{S}_\epsilon \mid  \frac{\bm{a}^T\x}{||\bm{a}||\cdot||\x||} \geq \alpha \}$ is defined as the aligned area on the sphere.
\end{definition}
\begin{definition}[Non-Aligned Area]
    The space $\mathcal{B}_{\theta} = \{\x \in \mathcal{S}_\epsilon \mid   \x\notin\mathcal{G}_{\alpha} \}$ is defined as the non-aligned area on the sphere.
\end{definition}
To elaborate, $\bm{a}$ is the rescaled vector of $q_n^{\prime} - d_n^+$ on the sphere.
The aligned area is defined as the set of embeddings whose cosine similarity with $\bm{a}$ is greater than $\alpha$, i.e., the hyper-spherical cap, while the non-aligned area is defined as the complement of the aligned area on the sphere.
The size of the cap is affected by the parameter $\alpha$, the greater $\alpha$, the smaller the size of the cap (i.e., the smaller aligned area).
To provide an intuitive understanding, we describe such a three-dimensional sphere as shown in Figure~\ref{fig: sphere}. 
With the above definition, Theorem~\ref{theo} describes the relation between dimensions and the probability of landing on the aligned area\footnote{The Theorem is proved in the literature~\cite{tkocz2012upper, becker2016new, ball1997elementary}}.
\begin{theorem}
\label{theo}
    Let $\mu$ be the Lebesgue measure\cite{halmos2013measure} on $\mathbb{R}^n$.
    The ratio of the measure of aligned area $\mathcal{G}_{\alpha}$ to the sphere $\mathcal{S}_\epsilon$ is subject to exponential decay  with respect to $n$:
    \begin{equation}
        \text{ratio} = \frac{\mu (\mathcal{G}_{\alpha})}{\mu (\mathcal{S}_\epsilon)} \leq \exp{(-n\alpha^2/2)}.
    \end{equation}
\end{theorem}
This Theorem~\ref{theo} indicates that the ratio of the measure on the aligned area to the whole hyper-sphere is at least exponentially small with respect to the dimension of the representation space.
Since the session query $q_n^{s}$ is distributed uniformly on the sphere $\mathcal{S}_\epsilon$, the probability of $q_n^{s}$ landing in the aligned area would be proportional to the ratio.
Considering the high dimensionality of existing PLM models, e.g., BERT~\cite{devlin2019bert} with 768 embedding sizes, the $q_n^{s}$ lands on the non-aligned area with high probability is inevitable without additional constraint. 
Fortunately, the available rewritten query $q_n^{\prime}$ with explicit search intent could help alleviate this situation.
To achieve query representation alignment by enabling $q_n^s$ to fall in the aligned area, a direct way is to minimize the distance between $q_n^s$ and $q_n^{\prime}$.
Because the defined anchor $\bm{a}$ is the center of the aligned area and the projection of $q_n^{\prime}$ on sphere $\mathcal{S}_{\epsilon}$,  $\bm{a}$ should have the smallest distance with  $q_n^{\prime}$ in the whole representation space $\mathcal{X}$.
Then, since the rewritten query $q_n^{\prime}$ is pre-encoded and $q^s_n$ always exists on the sphere, minimizing the distance between $q_n^s$ and $q_n^{\prime}$ for achieving query representation alignment is equivalent to bring $q_n^s$ closer to the anchor $\bm{a}$, i.e., fall into the aligned area.

\section{Methodology}
\subsection{Conversational Dense Retrieval}
A common practice for conversational dense retriever fine-tuning is to deploy the contrastive learning (CL) loss on the $n$-th session query $q^{s}_{n}$ with its relevant document $d_n^+$ and negatives $d_n^-$ encoded by an ad-hoc search retriever as Eq.~\ref{eq: CL}. The reformulated session query $q^{s}_{n}$ is the concatenation of the historical context, which is long and noisy, with the current query. The resulting query can hardly capture precisely the real search intent. 

\begin{equation}
\label{eq: CL}
    \mathcal{L}_{\text{CL}}(q^{s}_{n}, d_n^+, d_n^-) = -\log \frac{\exp\left(q^{s}_{n} \cdot d_n^+\right)}{\exp\left(q^{s}_{n} \cdot d_n^+\right) + \sum_{d_n^- \in D} \exp\left(q^{s}_{n} \cdot d_n^-\right)}
\end{equation}

\subsection{Query Representation Alignment} 
To enhance the conversational search ability, according to the aforementioned theoretical analysis (Section~\ref{sec: theory}), we can leverage the available rewritten queries (either automatic generation or manual annotations) to induce the learned session query representation in the ``alignment area''.
To achieve this, a straightforward and effective way is to employ the Mean Square Error (MSE) loss between the session query $q^{s}_{n}$ with both the rewritten query $q_n^{\prime}$ and the relevant document $d_n^+$.
Compared with only minimizing the similarity between the session query and the relevant document, deploying another MSE loss can establish the query representation alignment to enable the session query $q^{s}_{n}$ to approach both the relevant documents $d_n^+$ and the rewritten query $q^{\prime}_{n}$. 
Then, the two optimization objectives ensure neither $||q_n^s - d_n^+||^2 = 0$ nor $||q_n^s - q_n^{\prime}||^2 = 0$, 
but push the representation of $q_n^s$ to the balanced anchor point to achieve a better alignment effect. 
Besides, it enhances the robustness of the retriever, because of the scarcity of relevance judgment annotation in existing conversational search datasets, while there should be more relevant documents in real-world scenarios.
Such a base strategy is shown as Eq.~\ref{eq: two MSE}. Note that the rewritten query and relevant documents are only used as supervision signals in the training phase to obtain an aligned query representation, while not necessarily being used during the inference phase.

\begin{equation}
\label{eq: two MSE}
    \mathcal{L}_{\text{QRA}}^{\text{base}} = \text{MSE}\left(q^{s}_{n}, d_n^+\right) + \text{MSE}\left(q^{s}_{n}, q^{\prime}_{n}\right)
\end{equation}

\subsection{Using Hard Negatives} 
\label{subsec: usage of hard negatives}
The negatives play an important role in the training of dense retrieval models, especially the hard negatives~\cite{karpukhin2020dense}. Though effective, the utilization of negatives is lacking in the previous base strategy. To incorporate the hard negative, we further add one more MSE loss to move the session query representation $q^{s}_{n}$ away from the hard negative $\hat{d}_n^{-}$ as in Eq.~\ref{eq: two MSE with neg}. 
\begin{equation}
\label{eq: two MSE with neg}
    \mathcal{L}_{\text{QRA}}^{\text{neg.}} =  \text{MSE}\left(q^{s}_{n}, d_n^+\right) + \text{MSE}\left(q^{s}_{n}, q^{\prime}_{n}\right) - \text{MSE}\left(q^{s}_{n}, \hat{d}_n^{-}\right)
\end{equation}
This strategy shares a similar solution with the CL paradigm, which tries to enable the representation 
of query toward the relevant documents and keep away from the negatives during the training. 
However, the utilization of the hard negatives in MSE lacks an adaptive gradient update compared to CL. In terms of CL, the derivatives of $\mathcal{L}_{\text{CL}}$ with respect to $q_n^s$ is:
\begin{equation}
\frac{\partial \mathcal{L}_{\text{CL}}}{\partial q_n^s} =
\left(\frac{\exp (q_n^s \cdot \bm{d})}{\sum \exp (q_n^s \cdot \bm{d})} - 1 \right)d_n^+ + 
     \sum_{\bm{d}^-}\frac{\exp (q_n^s \cdot \bm{d})}{\sum \exp (q_n^s \cdot \bm{d})}d_n^-
\end{equation}
Thus, the coefficient $\frac{\exp (q_n^s \cdot \bm{d})}{\sum \exp (q_n^s \cdot \bm{d})}$ could dynamically adjust the weight on $d_n^-$, which enable the adaptive penalty on negative samples, i.e, if $q_n^s$ is close to $d_n^-$, the weight on $d_n^-$ would become large to keep them further away. 
To inherit this adaptive learning mechanism for hard negative, we incorporate the query representation alignment mechanism with the CL as Eq.~\ref{eq: two MSE with CL}. The contrastive paradigm combined strategy can also leverage the in-batch negative mechanism for potential benefit to improve the conversational dense retriever performance.
To further explore the combined impact of negatives in these strategies, we incorporate the query representation alignment and usage of (hard) negatives in both MSE and CL together as Eq.~\ref{eq: combine loss}. 

\begin{equation}
\label{eq: two MSE with CL}
\mathcal{L}_{\text{QRA}}^{\text{cont.}} = \mathcal{L}_{\text{QRA}}^{\text{base}} + \mathcal{L}_{\text{CL}}(q^{s}_{n}, d_n^+, d_n^-)
\end{equation}

\begin{equation}
\label{eq: combine loss}
\mathcal{L}_{\text{QRA}}^{\text{both}} = \mathcal{L}_{\text{QRA}}^{\text{neg.}} + \mathcal{L}_{\text{CL}}(q^{s}_{n}, d_n^+, d_n^-)
\end{equation}

\section{Experimental Setup}
\subsection{Datasets and Evaluation Metric}

\begin{table*}[t]
\centering
\caption{Statistics of datasets with different evaluation settings.}
\vspace{-3ex}
\resizebox{\textwidth}{!}{
\begin{tabular}{lccccccc|cccccc}
\toprule
\multirow{2}{*}{\textbf{Statistics}} & 
\multicolumn{2}{c}{\textbf{TopiOCQA}} & 
\multicolumn{2}{c}{\textbf{QReCC}} &
\textbf{CAsT-19} &  \textbf{CAsT-20} & \textbf{CAsT-21} & \multicolumn{2}{c}{\textbf{Diamond}} & \multicolumn{2}{c}{\textbf{Platinum}} & \multicolumn{2}{c}{\textbf{Gold}}\\
\cmidrule(lr){2-3}\cmidrule(lr){4-5}\cmidrule(lr){6-6}\cmidrule(lr){7-7}\cmidrule(lr){8-8}\cmidrule(lr){9-10}\cmidrule(lr){11-12}\cmidrule(lr){13-14}
 & \textbf{Train} &  \textbf{Test} & \textbf{Train} &  \textbf{Test} & \textbf{Test} & \textbf{Test} & \textbf{Test} & \textbf{Train} &  \textbf{Test} & \textbf{Train} &  \textbf{Test} & \textbf{Train} &  \textbf{Test}\\
\midrule
\# Conversations  & 3,509 & 205 & 10,823 & 2,775 & 20 & 25 & 18 & - & - & - & - & - & -\\
\# Turns (Queries) & 45,450 & 2,514 & 29,596 & 8,124 & 173 & 208 & 157 & 150,718 & 18,838 & 343,353 & 42,918 & 407,151 & 50,293 \\
\# Passages/Docs  & \multicolumn{2}{c}{25M} & \multicolumn{2}{c}{54M} & \multicolumn{2}{c}{38M} & 40M & \multicolumn{6}{c}{8.8M} \\ 
\bottomrule
\end{tabular}}
\label{table: dataset}
\vspace{-3ex}
\end{table*}

\noindent \textbf{\textit{Datasets}.} We use five conversational search datasets and an ad-hoc search dataset~\cite{arabzadeh2021matches} with three subsets separately to evaluate our methods in different settings. All datasets contain available rewritten queries for each original query and the statistic information is provided in Table~\ref{table: dataset}.
The TopiOCQA~\cite{adlakha2022topiocqa} and QReCC~\cite{anantha2021open} datasets are used to train conversational dense retrievers for the usual training-test evaluation. The TopiOCQA contains complex topic-shift phenomena within its conversation sessions, while the construction of QReCC focuses more on query rewriting.
The three\footnote{The CAsT-22 test set~\cite{owoicho2022trec} is designed for mixed-initiative conversational search, which is not appropriate used for our evaluation scenario.} CAsT datasets~\cite{dalton2020trec,dalton2021cast,dalton2022cast} with only a few samples are used for both out-of-domain zero-shot evaluation (i.e., directly apply the models trained on QReCC to the CAsT datasets) and in-domain cross-evaluation (i.e., train models on two of the CAsT datasets and evaluate on the remaining CAsT dataset).
The three subsets in~\cite{arabzadeh2021matches} are used to explore our methods in the ad-hoc search scenario. 

\noindent \textbf{\textit{Evaluation}.} For a thorough comparison with existing systems, we use four standard metrics: MRR, NDCG@3, Recall@10, and Recall@100 to evaluate the retrieval effectiveness.
We adopt the \textit{pytrec\_eval} tool~\cite{sigir18_pytrec_eval} for metric computation.

\subsection{Baselines}
We compare our method with two lines of conversational search approaches. The first line performs conversational query rewriting (CQR) based on generative rewriter models and off-the-shelf retrievers, including
(1) \textbf{QuReTeC}~\cite{voskarides2020query}: A weakly-supervised method to train a sequence tagger to decide whether each term contained in historical context should be used to expand the current query. 
(2) \textbf{T5QR}~\cite{lin2020conversational}: A strong T5-based model for query reformulation. 
(3) \textbf{CONQRR}~\cite{wu2022conqrr}: A T5-based model applying reinforcement-learning for query reformulation. 
(4) \textbf{ConvGQR}~\cite{mo2023convgqr}: Combining two T5-based models for query rewrite and query expansion in query reformulation. 
(5) \textbf{EDIRCS}~\cite{mao2023search}: A text-editing method to construct the search-oriented conversational query. 
(6) \textbf{IterCQR}~\cite{jang2023itercqr}: An iterative conversational query reformulation associated with IR signals feedback. 
(7) \textbf{InstructorR}~\cite{jin2023instructor}: An unsupervised conversational dense retrieval method instructed by LLM. 
(8) \textbf{LLM-Aided IQR}~\cite{ye2023enhancing}: An informative query rewriting method aided by prompting LLM. 
(9) \textbf{Human-Rewritten}~\cite{anantha2021open}: Manual annotations provided in original datasets.
The second line conducts conversational dense retrieval (CDR) fine-tuning based on ad-hoc search dense retrievers to learn the representation of the session query, including 
(10) \textbf{CQE-sparse}~\cite{lin2021contextualized}: A weakly-supervised method to expand important tokens from the context via contextualized query embeddings. 

The second line performs conversational dense retrieval (CDR), which are initialized from an ad-hoc search retriever and fine-tunes with conversational search data. 
(11) \textbf{ConvDR}~\cite{yu2021few}: A conversational dense retriever fine-tuned from ANCE by mimicking the representations of human rewrites, which has a similar alignment goal and is a main competitor with our methods. However, as discussed in Section~\ref{sec: Related Work}, this works does not consider the alignment between the session query and the gold passage (nor the misalignment with the negative passages).
(12) \textbf{SDRConv}~\cite{kim2022saving} ANCE fine-tuned on conversational search data with additionally mined hard negatives. 
(13) \textbf{Conv-ANCE}~\cite{mao2023learning}: ANCE fine-tuned on conversational search data only using the contrastive loss, which is another main competitor to our methods. 
(14) \textbf{HAConvDR}~\cite{mo2024history}: ANCE fine-tuned on context-denoising reformulated query and additional signals from historical turns based on the impact of retrieval effectiveness.
(15) \textbf{LeCoRe}~\cite{mao2023learning}: The current state-of-the-art method, which extends SPLADE with two well-matched multi-level denoising methods, which cannot be fairly compared with the other methods.\footnote{All compared models are initialized from ANCE except LeCoRe is from SPLADE. Since TopiOCQA and QReCC are open benchmarks, we compare with the performance reported in the corresponding baseline papers.}

\subsection{Implementation Details}
We implement the conversational dense retriever based on ANCE~\cite{xiong2020approximate} using the PyTorch and Huggingface libraries. The conversational dense retrievers are fine-tuned with each corresponding training strategy $\mathcal{L}_{\text{QRA}}$. 
Following previous studies~\cite{yu2021few,mao2023learning,mo2023learning}, only the session query encoder will be trained while the encoders for rewritten queries and documents are frozen. The lengths of the query turn, session query, and document are truncated into 64, 512, and 384, respectively, to fit the majority of examples in the dataset. The batch size is set to 32 in accordance to our computational resources. We use Adam optimizer with 1e-5 learning rate and set the training epoch to 10. The dense retrieval is performed using Faiss~\citep{johnson2019billion}. More details can be found in our released code.\footnote{\url{https://github.com/fengranMark/QRACDR}}

\section{Results}
\begin{table*}[t]
    \centering
    \caption{Performance of different dense retrieval methods on two datasets. $\dagger$ denotes significant improvements with t-test at $p<0.05$ over each of the compared CDR systems. \textbf{Bold} and \underline{underline} indicate the best and the second-best results.}
    \vspace{-2ex}
    \begin{tabular}{c|l|cccc|cccc}
    \toprule
    \multirow{2}{*}{Category} & \multicolumn{1}{c|}{\multirow{2}{*}{{Method}}} & \multicolumn{4}{c|}{TopiOCQA} &
    \multicolumn{4}{c}{QReCC}\\
    \cmidrule(lr){3-6} \cmidrule(lr){7-10}
     & & {MRR} & {NDCG@3} & {Recall@10} & {Recall@100} & {MRR} & {NDCG@3} & {Recall@10} & {Recall@100} \\
    \midrule
    \multirow{8}{*}{CQR} & QuReTeC & 11.2 & 10.5 & 20.2 & 34.4 & 35.0 & 32.6 & 55.0 & 72.9\\
    & T5QR & 23.0 & 22.2 & 37.6 & 54.4 & 34.5 & 31.8 & 53.1 & 72.8\\
    & CONQRR & - & - & - & - & 41.8 & - & 65.1 & 84.7\\
    & ConvGQR & 25.6 & 24.3 & 41.8 & 58.8 & 42.0 & 39.1 & 63.5 & 81.8 \\
    & EDIRCS & - & - & - & - & 42.1 & - & 65.6 & 85.3\\
    & IterCQR & 26.3 & 25.1 & 42.6 & 62.0 & 42.9 & 40.2 & 65.5 & 84.1\\ 
    & LLM-Aided IQR & - & - & - & - & 43.9 & 41.3 & 65.6 & 79.6\\
    & Human-Rewritten & - & - & - & - & 38.4 & 35.6 & 58.6 & 78.1\\
    \midrule 
    \multirow{11}{*}{CDR} &  CQE-sparse & 14.3 & 13.6 & 24.8 & 36.7 & 32.0 & 30.1 & 51.3 & 70.9\\
    & InstructorR & 25.3 & 23.7 & 45.1 & 69.0 & 43.5 & 40.5 & 66.7 & 85.6\\
    & SDRConv & 26.1 & 25.4 & 44.4 & 63.2 & 47.3 & 43.6 & 69.8 & 88.4 \\
    & Conv-ANCE & 22.9 & 20.5 & 43.0 & 71.0 & 47.1 & 45.6 & 71.5 & 87.2\\
    & ConvDR & 27.2 & 26.4 & 43.5 & 61.1 & 38.5 & 35.7 & 58.2 & 77.8\\
    & HAConvDR & 30.1 & 28.5 & 50.8 & 72.8 & 48.5 & 45.6 & 72.4 & \underline{88.9} \\
    & LeCoRe (SPLADE) & 32.0 & 31.4 & 54.3 & 73.5 & \underline{51.1} & \underline{48.5} & 73.9 & \textbf{89.7}\\
    \cmidrule(lr){2-10}
    & Our QRACDR & 31.8 & 30.6 & 50.0 & 67.6 & 47.1 & 44.5 & 71.4 & 87.1\\
    & \quad + negative & 32.4$^\dagger$ & 31.3 & 51.5 & 71.4 & 48.5 & 45.8 & 72.1 & 88.1\\
    & \quad + contrastive & \textbf{37.7}$^\dagger$ & \textbf{36.5}$^\dagger$ & \textbf{57.1}$^\dagger$ & \textbf{75.8}$^\dagger$ & \textbf{51.6}$^\dagger$ & \textbf{49.1}$^\dagger$ & \textbf{74.8}$^\dagger$ & \textbf{89.7}\\
    & \quad + both & \underline{34.6}$^\dagger$ & \underline{33.5}$^\dagger$ & \underline{54.4} & \underline{74.0}$^\dagger$ & 50.9 & 48.3 & \underline{74.1} & \underline{88.9}\\ 
    \bottomrule
    \end{tabular}
    \label{table: Main Results}
\vspace{-2ex}
\end{table*}

\begin{table*}[t]
    \centering
    \caption{Dense retrieval performance on three CAsT datasets. Out-of-domain (OOD) denotes zero-shot evaluation by directly applying the models trained on QReCC to the CAsT datasets. In-domain (ID) denotes cross-evaluation by training models on two of the CAsT datasets and evaluating on the remaining CAsT dataset.
    All the compared systems are tested with the OOD setting. 
    $\dagger$ denotes significant improvements with t-test at $p<0.05$ over the main competitors, Conv-ANCE and ConvDR. \textbf{Bold} and \underline{underline} indicate the best and the second-best results.}
    \vspace{-2ex}
    \begin{tabular}{l|ccc|ccc|ccc}
    \toprule
        \multicolumn{1}{c|}{\multirow{2}{*}{Method}} & \multicolumn{3}{c|}{CAsT-19} & \multicolumn{3}{c|}{CAsT-20} & \multicolumn{3}{c}{CAsT-21} \\ 
        \cmidrule(lr){2-4} \cmidrule(lr){5-7} \cmidrule(lr){8-10} 
        & MRR     & NDCG@3   & Recall@100  & MRR     & NDCG@3   & Recall@100  & MRR     & NDCG@3   & Recall@100  \\ 
        \midrule
        QuReTeC & 68.9 & 43.0 & 33.7 & 43.0 & 28.7 & 34.6 & - & - & -\\
        T5QR & 70.8 & 42.6 & 33.2 & 42.8 & 30.7 & 35.3 & 36.3 & 24.9 & 29.0 \\
        ConvGQR & 70.8 & 43.4 & 33.6 & \textbf{46.5} & \textbf{33.1} & \underline{36.8} & 43.3 & 27.3 & 33.0\\
        CQE-sparse & 67.1 & 40.9 & 33.5 & 42.3 & 28.9 & 35.6 & - & - & -\\
        InstructorR & 61.2 & \textbf{46.6} & \underline{34.4} & 43.7 & 29.6 & \textbf{40.8} & - & - & - \\
        Conv-ANCE & 66.2 & 40.1 & 29.2 & 42.5 & 26.5 & 31.6 & 36.3 & 23.5 & 34.3\\
        ConvDR & \underline{71.7} & 43.9 & 32.2 & 43.9 & \underline{32.4} & 33.8 & 45.6 & \textbf{36.1} & \textbf{37.6} \\
        \midrule
        Our QRACDR (OOD) & 71.2 & 44.7$^\dagger$ & 34.1$^\dagger$ & 42.7 & 29.2 & 32.2 & 41.6 & 29.0 & 35.2 \\
        \quad + negative & \textbf{73.1}$^\dagger$ & \underline{45.1}$^\dagger$ & \textbf{35.2}$^\dagger$ & \underline{44.2}$^\dagger$ & 30.3 & 32.4 & \textbf{47.1}$^\dagger$ & \underline{31.3} & \textbf{37.6}\\
        \quad + contrastive & 65.1 & 38.4 & 32.6 & 41.5 & 27.1 & 33.7 & \underline{46.5} & 30.4 & 35.9\\ 
        \quad + both & 64.5 & 37.6 & 32.8 & 39.5 & 25.6 & 33.3 & 44.9 & 29.4 & \underline{36.0}\\
        \midrule
        Our QRACDR (ID) & 63.7 & 39.9 & 29.7 & 38.0 & 25.8 & 27.4 & 41.5 & 26.2 & 32.6\\
        \quad + negative & 64.6 & 40.4 & 30.8 & 38.4 & 25.9 & 27.1 & 43.4 & 28.2 & 31.5\\
        \quad + contrastive & 61.1 & 36.3 & 27.8 & 36.6 & 22.0 & 28.6 & 43.0 & 28.0 & 32.3\\ 
        \quad + both & 58.1 & 33.6 & 26.6 & 34.6 & 21.4 & 27.0 & 38.8 & 22.7 & 28.9\\
        \bottomrule
     \end{tabular}
     \label{table: Zero-shot}
\vspace{-2ex}
\end{table*}

\begin{table*}[!t]
\centering
\caption{Effect of incorporating rewritten queries and relevant documents via our base QRACDR on five datasets.}
\vspace{-3ex}
\begin{tabular}{l|cc|cc|cc|cc|cc}
\toprule
  & \multicolumn{2}{c|}{TopiOCQA} & \multicolumn{2}{c|}{QReCC} & \multicolumn{2}{c|}{CAsT-19} & \multicolumn{2}{c|}{CAsT-20} & \multicolumn{2}{c}{CAsT-21}\\
\cmidrule(lr){2-3}\cmidrule(lr){4-5}\cmidrule(lr){6-7}\cmidrule(lr){8-9}\cmidrule(lr){10-11}
~ & {MRR} & {NDCG@3} & {MRR} & {NDCG@3} & {MRR} & {NDCG@3} & {MRR} & {NDCG@3} & {MRR} & {NDCG@3}\\
\midrule
Base QRACDR (Eq.~\ref{eq: two MSE}) & 31.8 & 30.6 & 47.1 & 44.5 & 73.1 & 45.1 & 42.7 & 29.3 & 41.6 & 29.0\\
\quad w/o $\text{MSE}\left(q^{s}_{n}, q^{\prime}_{n}\right)$ & 30.8 & 29.4 & 40.8 & 38.4 & 59.2 & 35.3 & 34.2 & 22.8 & 39.9 & 25.1 \\
\quad w/o $\text{MSE}\left(q^{s}_{n}, d_n^+\right)$ & 24.7 & 23.7 & 41.5 & 38.5 & 64.9 & 39.6 & 34.9 & 21.2 & 40.0 & 26.2\\
\bottomrule
\end{tabular}
\label{table: ablation_different_information}
\vspace{-3ex}
\end{table*}

\subsection{Evaluation Results}
\label{sec: Normal Evaluation}
We first conduct the usual training-testing evaluation on TopiOCQA and QReCC with different training strategies. The overall performance is shown in Table~\ref{table: Main Results}. We can make the following observations: 

(1) The best variants of our QRACDR consistently outperform all compared methods across four metrics on two datasets, demonstrating superior effectiveness on conversational search. We observe that our best results have 16.2\% and 1.2\% NDCG@3 relative gains on TopiOCQA and QReCC over the state-of-the-art LeCoRe (based on SPLADE) and 38.3\% and 7.7\% NDCG@3 gains over the second-best compared system with the same retriever (based on ANCE).
The superior effectiveness can be attributed to two different aspects. (i) Making the session query representation closer to both the representation of rewritten queries and relevant documents can better help the fine-tuned conversational dense retriever address user search intent implicitly. (ii) Incorporating negative samples with suitable mechanisms, e.g., adaptive negative learning in CL, helps achieve better performance (than ConvDR and Conv-ANCE).

(2) Among different training strategies for our QRACDR, the base model can still surpass all compared systems on TopiOCQA whose retrievers are the same and achieve comparable results on QReCC, showing the simplicity and effectiveness of query representation alignment. 
Besides, further adding hard negatives can improve the search results, and employing the contrastive paradigm with negatives is the best variant. Such improvements indicate the importance of incorporating negatives in dense retriever training. However, simultaneously incorporating hard negatives in MSE and CL loss functions would damage the model performance. This might be because the effectiveness of leveraging negatives is fully exploited by CL, which has a better adaptive mechanism as discussed in Sec.~\ref{subsec: usage of hard negatives}. 

(3) Comparing the CQR and CDR methods, we see that CDR systems generally have better performance (e.g. ConvDR, Conv-ANCE, and LeCoRe). This might be because CQR is trained to optimize query rewriting only, without taking into account the quality of search results.
However, incorporating the rewrites produced by CQR into CDR, such as ConvDR, LeCoRe, and our QRACDR, can further improve the retrieval performance, suggesting that combining CDR and CQR is beneficial.
This is also the intuitive motivation for designing our methods, i.e., making query representation alignment with both rewritten query and relevant document. 
Besides, we observe that LLM-based CQR systems (IterCQR and LLM-Aided IQR) and CDR systems (InstructorR) cannot outperform existing state-of-the-art CDR systems. We still need to find a better way to leverage LLM, possibly in connection with conversational search.

\subsection{Low-resource Evaluation}
We then conduct low-resource testing on three CAsT datasets to evaluate the transferring abilities of our QRACDR and explore the impact of data distribution. The results are reported in Table~\ref{table: Zero-shot}, which shows the following observations: 

(1) The generalizability of each variant of our QRACDR is different, and the base QRACDR employed with hard negative performs the best in both the OOD and ID scenarios. Among OOD settings, our best approach outperforms the compared systems on most metrics and achieves comparable results on the remaining, which confirms the transferring abilities of QRACDR. 

(2) The QRACDR models trained with contrastive loss do not perform well on top-rank results, implying that it might not be necessary to incorporate too many mechanisms for the desired transferring ability, though they might perform better on recall.

(3) The models in OOD perform better than the ones in ID, which might be attributed to two aspects. (i) Though the in-domain training data might have better data distribution, the samples are still not enough to achieve better retrieval performance, compared with the models trained with much more out-of-domain conversational search data. (ii) The data distribution gap between each dataset might not fit with our intuition. For example, the data distribution similarity between QReCC with each CAsT dataset might be larger than that among the CAsT dataset, since each CAsT dataset is constructed with different information-seeking goals and the topic of conversation sessions varies from each year. Thus, the large-scale conversational search data are valuable for building transferable conversational retrieval models.

\begin{table*}[t]
    \centering
    \caption{Effect of incorporating our QRACDR approach on three ad-hoc search subsets with available rewritten queries.}
    \vspace{-2ex}
    \begin{tabular}{l|ccc|ccc|ccc}
    \toprule
        \multicolumn{1}{c|}{\multirow{2}{*}{Method}} & \multicolumn{3}{c|}{Diamond} & \multicolumn{3}{c|}{Platinum} & \multicolumn{3}{c}{Gold} \\ 
        \cmidrule(lr){2-4} \cmidrule(lr){5-7} \cmidrule(lr){8-10} 
        & MRR & NDCG@3 & Recall@10 & MRR & NDCG@3 & Recall@10 & MRR  & NDCG@3 & Recall@10  \\ 
        \midrule
        Our QRACDR & 25.69 & 23.14 & 47.79 & 27.92 & 24.40 & 53.90 & 26.03 & 23.02 & 47.80 \\
        \quad w/o QRA & 23.90 & 21.23 & 46.78 & 19.21 & 15.79 & 39.69 & 23.08 & 19.99 & 43.96 \\
        \quad w/o Fine-tuning & 23.37 & 20.64 & 41.95 & 17.79 & 14.23 & 33.80 & 21.83 & 18.65 & 41.81 \\
        \bottomrule
     \end{tabular}
     \label{table: ad-hoc results}
\vspace{-3ex}
\end{table*}

\subsection{Ablation Studies}
In this section, we first conduct ablation studies to investigate the effects of incorporating various features via our base QRACDR. Then, we explore the impacts of different decomposition terms in combining MSE and CL loss functions, which is the best variant of QRACDR in supervised evaluation.
\subsubsection{Supervision effectiveness}
Table~\ref{table: ablation_different_information} shows the effectiveness of leveraging rewritten queries and relevant documents in our base QRACDR (Eq.~\ref{eq: two MSE}) on five datasets. 
We observe that removing any information leads to performance degradation, indicating the usefulness of training the model with each information. 
The utilization of rewritten queries is more crucial than the relevant documents, except in TopiOCQA. 
This might be attributed to TopiOCQA having lower-quality rewritten queries, which have been automatically generated.
Besides, the results show that the retriever could remain relatively effective when trained without relevant document supervision. This observation indicates that the rewritten queries could indeed provide useful signals for aligning with the search intent. 

\begin{figure}[t]
    \centering  \includegraphics[width=\linewidth]{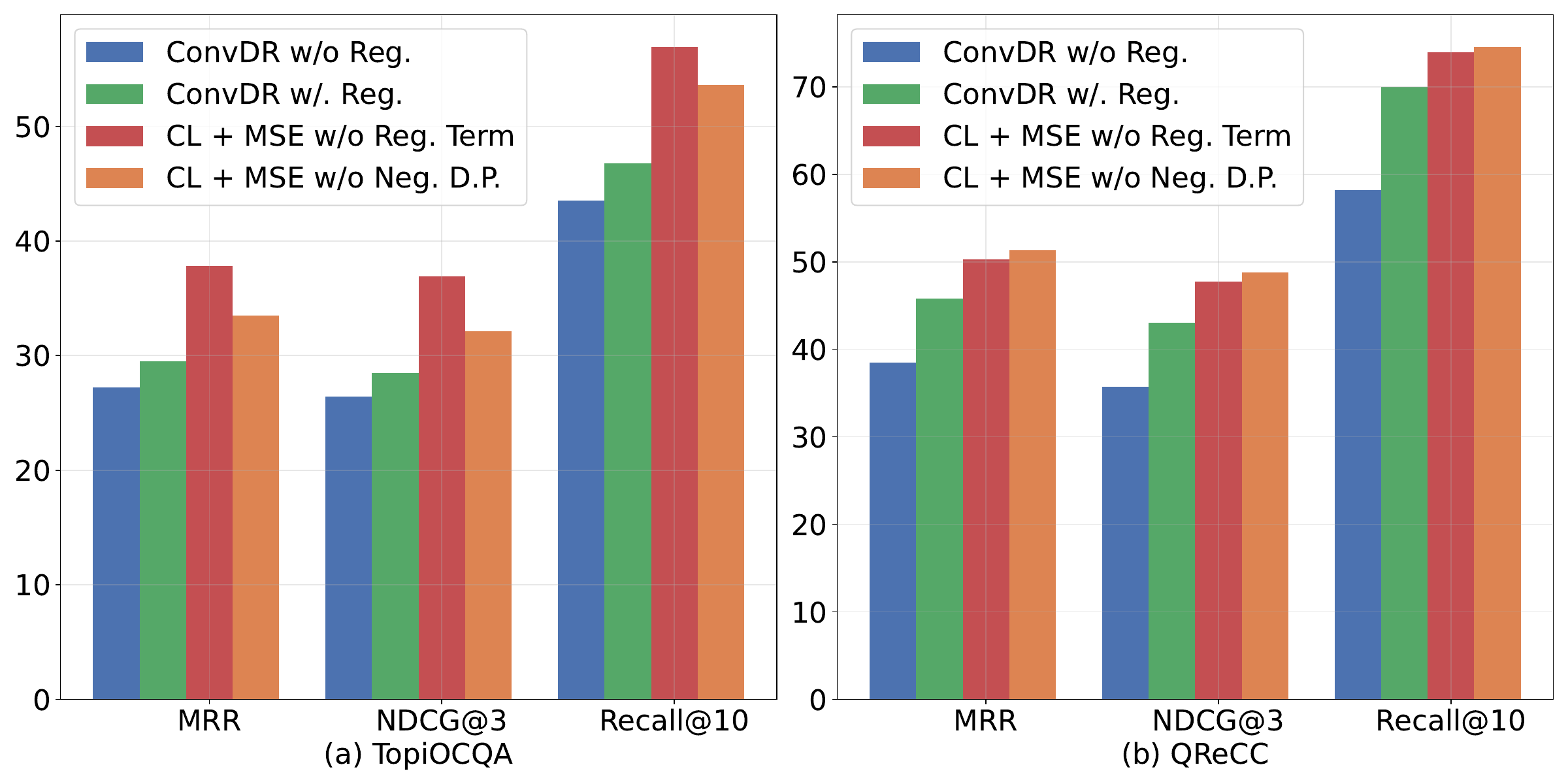}
    \vspace{-3ex}
    \caption{Impacts of decomposition terms in MSE based on our best QRACDR (CL + MSE) and regularization in ConvDR.}
    \label{fig: Term in MSE}
\vspace{-3ex}
\end{figure}

\subsubsection{Impacts of Combining MSE and CL}
The best variant of our QRACDR (Eq.~\ref{eq: two MSE with CL}) for normal evaluation is to combine the MSE function and the CL function.
Compared with commonly used CL, the core of QRACDR is the additional two MSE functions aiming for query representation alignment. 
To better understand the impact of the mechanisms behind, we first rewrite the two deployed MSE functions into two terms, namely, a regularized term and a negative dot product term as in Eq.~\ref{eq: decomposed term}. 
Thus, the $v$ could be either $d_n^{+}$ or $q_n^{\prime}$ in Eq.~\ref{eq: two MSE}. Then, we analyze these two decomposition terms by separately incorporating them with CL.
Besides, the compared system ConvDR with a similar use of rewritten query supervision lacks regularization in calculating query-document similarity. Thus, we also investigate the effect of regularization under the setting of conversational dense retrieval (ConvDR).
\begin{equation}
\label{eq: decomposed term}
    \text{MSE}(q_n^s, v) = ||q_n^s - v||^2 = \underbrace{||q_n^s||^2 + ||v||^2}_{\text{regularized term}} - \underbrace{2 \cdot q_n^s \cdot v}_{\text{neg. dot product}}
\end{equation}

The results are shown in Figure~\ref{fig: Term in MSE}, where we find that adding regularization improves the performance of ConvDR, which is more significant on QReCC.
However, it still cannot surpass the performance of our QRACDR, which demonstrates our approaches can better address the search intent. Within the decomposition terms in MSE, the regularization contributes more to QReCC, while the negative dot product shows higher importance on TopiOCQA. 
Intuitively, considering that the conversation sessions are more complex and challenging in TopiOCQA, the negative dot product mechanism should be more important to enable session query representation $q_n^s$ to become closer to the aligned area, while the regularization term prevents the over-fitting on the relatively easy QReCC.

\subsection{Ad-hoc Search Scenarios Analysis}
In this section, we explore our method to understand whether leveraging available rewritten queries is also effective in ad-hoc search via alignment of query representation.

Table~\ref{table: ad-hoc results} shows the comparison among models using QRACDR (Eq.~\ref{eq: two MSE with CL}), the common CL without query representation alignment (Eq.~\ref{eq: CL}), and without further fine-tuning in three subsets with different types of rewritten queries~\cite{arabzadeh2021matches}. The quality of both original and rewritten queries decreases with the order: Diamond, Platinum, and Gold, where the Diamond set has the best rewritten queries.
Overall, our QRACDR achieves various degrees of improvement across all metrics, which demonstrates the effectiveness of the query representation alignment mechanism. 
The improvement is much larger in the Platinum and Gold than in Diamond. This may suggest that our method based on query representation alignment is particularly effective in improving the original queries of lower quality. 

\subsection{Learned Query Representations Analysis}
In this section, we analyze the model behavior with different training strategies according to their learned representation. To show the effect of query representation alignment, the comparisons are conducted among the variants of our QRACDR and a main competitor ConvDR, without specific query representation alignment. This allows us to fully attribute the difference in performance to different alignment strategies.

\begin{figure}[t]
    \centering  \includegraphics[width=\linewidth]{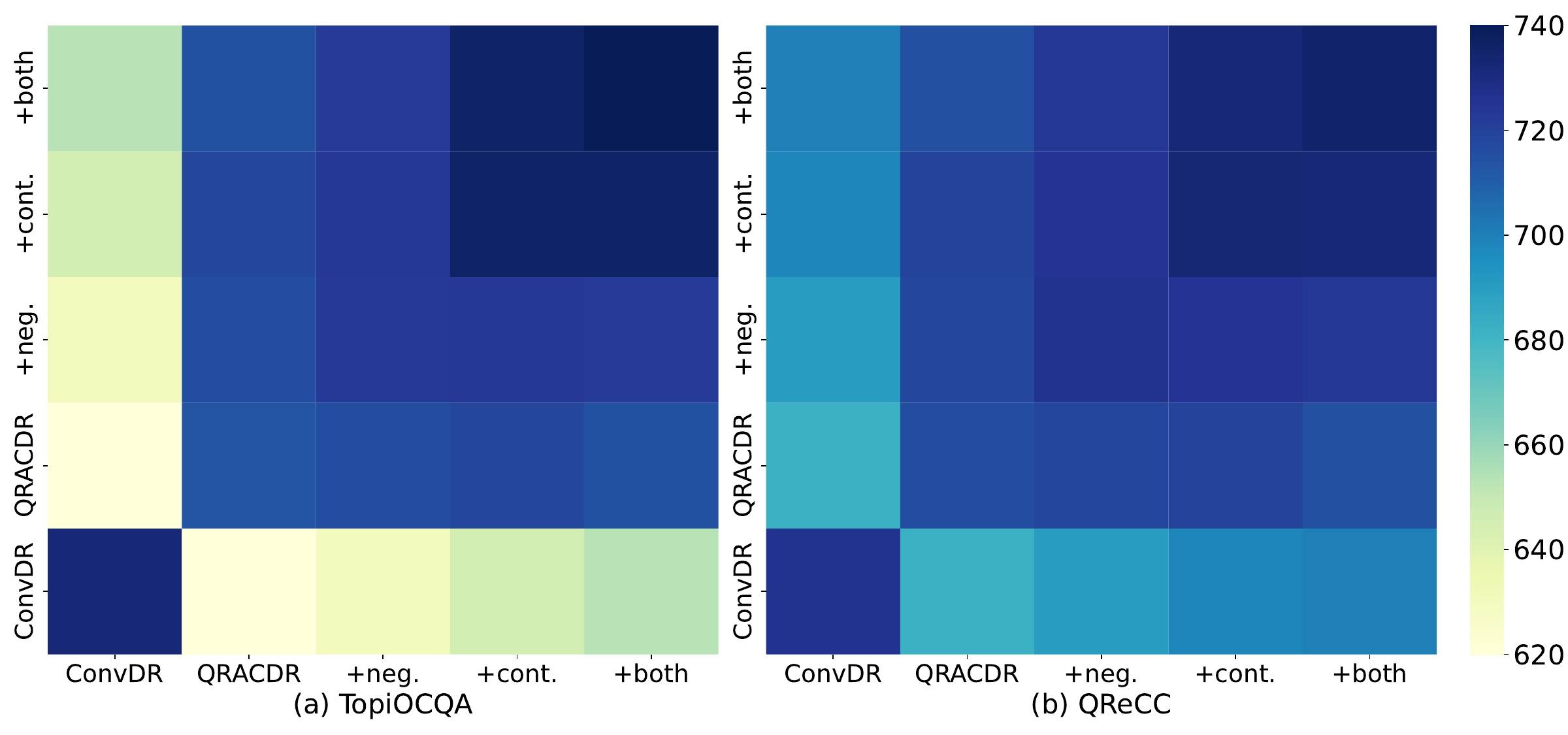}
    \vspace{-3ex}
    \caption{Model behavior at the average dot product score of two representations of the same query on two datasets.}
    \label{fig: heatmap_q-q}
\vspace{-3ex}
\end{figure}

\noindent \textbf{Q-Q Similarities.}
We first analyze the average dot product similarity between the representation of the same query learned by different models on two datasets, which is shown in Figure~\ref{fig: heatmap_q-q}. We can find that the query representation of ConvDR is very different from our QRACDR. This shows directly the effect of applying our alignment techniques. The representation differences are more distinct on TopiOCQA than on QReCC, showing a more pronounced impact of query representation alignment in TopiOCQA for longer and more complex topic-shift conversations. 
Among our proposed approaches, the incorporation of contrastive loss or negatives also affects the similarity.

\begin{figure}[t]
    \centering  \includegraphics[width=\linewidth]{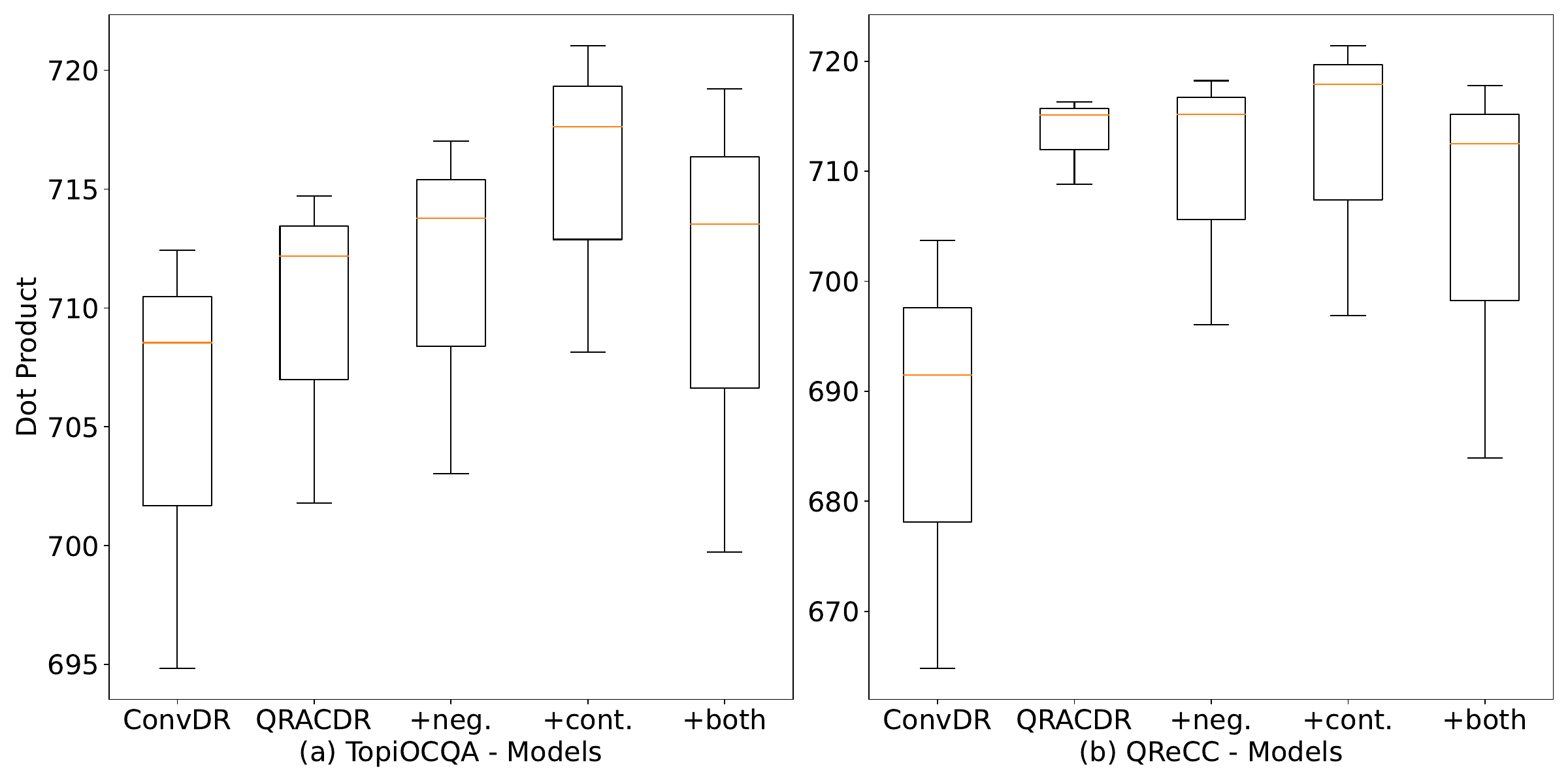}
    \vspace{-3ex}
    \caption{Model behavior at the average dot product score of queries and their nearest relevant documents on two datasets.}
    \label{fig: boxplot_q-d}
\vspace{-3ex}
\end{figure}

\noindent \textbf{Q-D Similarities.}
The effects on query representation also influence their similarities with relevant documents. 
As shown in Figure~\ref{fig: boxplot_q-d}, the variant of our QRACDR incorporated with contrastive paradigm has the highest similarity on both datasets, which is consistent with the previous main evaluation of performance (Table~\ref{table: Main Results}). 
Besides, the gap in the results is much more significant on TopiOCQA and the main competitor ConvDR is still lower than our proposed methods, again, confirming the effectiveness of query representation alignment on lengthy and complex conversations.

\subsection{Impact of Context}
\begin{figure}[t]
    \centering  \includegraphics[width=\linewidth]{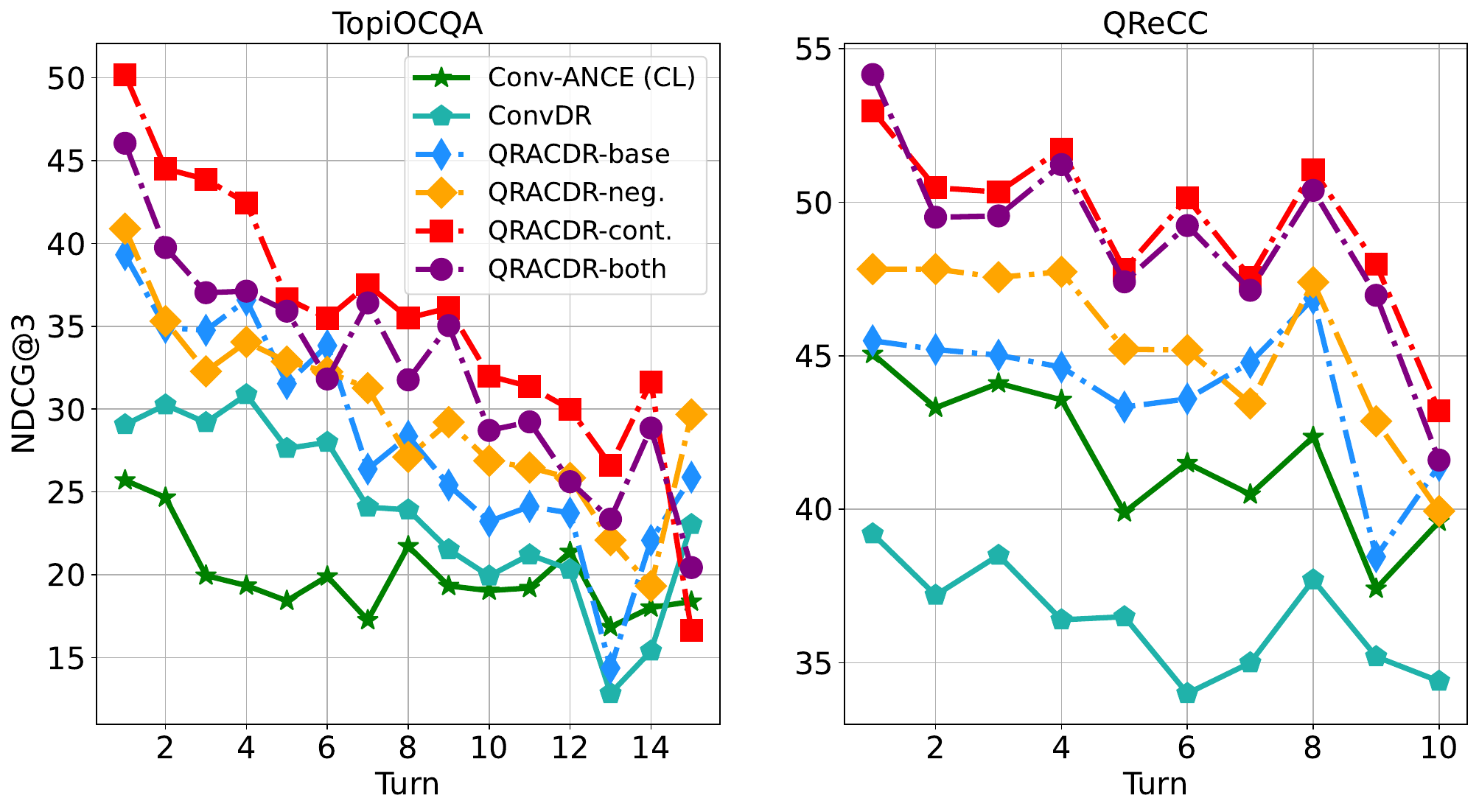}
    \vspace{-3ex}
    \caption{Model behavior at the NDCG@3 score with different conversation turns on two different datasets.}
    \label{fig: turn-ndcg}
\vspace{-3ex}
\end{figure}

In this experiment, we study the impact of the context (multi-turn conversations) for the query representation learned by different models. We use their per-turn retrieval performance for evaluation.
As shown in Figure~\ref{fig: turn-ndcg}, the NDCG@3 of QRACDR variants drops as the conversation goes on, similar to the existing systems (Conv-ANCE and ConvDR). This is because the queries in later conversation turns are more likely to depend on previous turns, while the topic-shift and long-tail phenomenon makes the context dependency harder to resolve.
Nevertheless, our QRACDR maintains its performance above Conv-ANCE and ConvDR throughout the conversation turns, indicating our query representation alignment technique helps the retrieval model to capture the context information required to understand the user’s information needs.

\section{Conclusion}
In this paper, we propose a method for conversational dense retriever exploiting the idea of query representation alignment, QRACDR. It leverages both the rewritten query and the relevant document to better determine the user search intent. 
Several training strategies are proposed to incorporate various information in dense retrieval. Experimental results on eight datasets with different settings demonstrate the effectiveness of our methods. The detailed analysis provides an understanding of the behavior of the models and confirms the effectiveness of query representation alignment.
In the future, we plan to explore better ways of integrating the advantages of the CQR and CDR methods, i.e., a more effective way to fuse the representation of rewritten queries and the candidate documents, to further improve the performance of conversational search.
\begin{acks}
This work is supported by a discovery grant from the Natural Science and Engineering Research Council of Canada and a Talent Fund of Beijing Jiaotong University (2024JBRC005).
\end{acks}

\bibliographystyle{ACM-Reference-Format}
\bibliography{sample-base}

\appendix


\end{document}